\documentclass[a4paper,12pt]{article}
\makeatletter
\@addtoreset{equation}{section}
\makeatother

\def\greek2{I\hspace{-.1em}I}
\def\beeqno{\begin{eqnarray*}}
\def\eneqno{\end{eqnarray*}}
\def\beeq{\begin{eqnarray}}
\def\eneq{\end{eqnarray}}
\def\bear{\begin{array}}
\def\enar{\end{array}}
\def\*{exp\{i\frac{\theta^{\mu\nu}}{2}\partial_\mu^\alpha \partial_\nu^\beta\}}
\def\fra12{\frac{1}{2}}

\def\hsp1{\hspace{1cm}}

\def\d{\delta}
\def\D{\Delta}
\def\g{\gamma}

\def\k{\kappa}

\def\L{\Lambda}
\def\m{\mu}
\def\n{\nu}

\def\r{\rho}
\def\s{\sigma}

\def\e{\eta}

\def\part{\partial}

\def\={& \hspace{-.3cm}= \hspace{-.3cm}&}
\def\:={& \hspace{-.3cm}:= \hspace{-.3cm}&}

\begin{document}

\begin{titlepage}

\vspace{3cm}
\begin{flushright}
CHIBA-EP-138
\end{flushright}

\vspace{1cm}

\begin{center}
{\Large \bf
Uncertainty of Velocity in $\k$-Minkowski Spacetime}\\

\vspace{2cm}

Ken-Ichi Tezuka\footnote{e-mail: tezuka@graduate.chiba-u.jp}  \\

\vspace{1cm}

Graduate School of Science and Technology, Chiba University, Japan \\

\vspace{5cm}

\end{center}
\begin{abstract}
A velocity of a point particle in the $\k$-Minkowski spacetime is investigated. 
Characteristic points of the spacetime are that the Poincare group becomes a 
quantum group with $\k$, which is a mass dimension parameter, and is a 
kind of non-commutative geometry. We consider a particle in a
coordinate space instead of it in a momentum space which is discussed in many articles. 
We see that the particle's velocity has an uncertainty which depends on a length of  
particle's propagation. 
\end{abstract}

\end{titlepage}
\section{Introduction}
In cosmic ray physics, the following problems are not solved yet; anomalous 
detection of extremely high 
energy cosmic rays \cite{Takeda:1998ps} above the GZK cutoff ($5 \times 10^{19}$ eV) 
\cite{greisen}, and TeV-$\g$ rays \cite{Quinn:dj} from Mrk 501 and Mrk 421. It was 
suggested that the absence of 
the GZK cutoff is closely related to violation of the Lorentz invariance \cite{sato}. If 
the Lorentz 
invariance is violated or deformed, there is the possibility to explain that there is not 
the GZK cutoff at $E=5 \times 10^{19}$ eV \cite{Coleman:1998ti,tamaki}.
Therefore we have to consider a mechanism for the violation or deformation of the 
symmetry. 
\par 
One of 
the characteristic features with respect to quantum gravity is the Plack length. If 
we would like to deal with this as an observer independent scale, the special relativity 
becomes a deformed one which has two scales; a velocity $c$ and a mass $\k$. This 
is called the doubly special relativity \cite{Amelino-Camelia:2000mn} which 
is explicitly realized by the $\k$-Poincare algebra \cite{Kowalski-Glikman:2001gp}. 
\par
The $\k$-Poincare algebra is defined by replacing the Poincare algebra with a 
quantum group (Hopf algebra) with a parameter $\k$ \cite{Lukierski:1991pn}.  
In the limit $\k \to \infty$, the Hopf algebra is reduced to the ordinary Lie 
algebra. Majid and Ruegg \cite{Majid:1994cy} have pointed out that the $\k$-deformed 
algebra is the symmetry in the $\k$-Minkowski spacetime in which
time and spatial coordinates are non-commutative.
\par
Some authors have discussed particle's velocity formulae
\cite{Lukierski:1993wx,tamaki} in the $\k$-Minkowski spacetime. In the 
arguments, 
it was assumed that a velocity is defined by $v=\frac{\part E}{\part p}$ which is true 
in the commutative Minkowski spacetime. In \cite{Lukierski:1993wx}, by applying the 
formula to a massless particle, it was  predicted that the speed of light is dependent 
on its energy $E$ and the parameter $\k$.  
In \cite{tamaki}, by changing the rule of differentiation with respect to 
momenta, it was mentioned that the speed of light is undeformed. However validity of using
the formula $v=\frac{\part E}{\part p}$
in the $\k$-Minkowski spacetime has not been discussed sufficiently yet. 
\par
It is known that there exists some bases of the $\k$-Poincare algebra
\cite{Lukierski:1993wx,Magueijo:2001cr}, which are related 
with each other by redefinition of the translation operators 
\cite{Kowalski-Glikman:2002we}. In other words, the $\k$-deformation of the Poincare 
algebra is not unique. Since the dispersion relation is given as the Casimir operator 
of the algebra, the 
particle's velocity depends which bases we use, if the formula 
$v=\frac{\part E}{\part p}$ is used. In this paper we would like to consider a 
velocity by avoiding using the formula, and would like to compare results.
\par
The present paper is organized as follows. In the next section, we review the 
$\k$-Minkowski 
spacetime and the $\k$-Poincare algebra. Some velocity formulae are given explicitly by 
using the bicrossproduct basis. In the section 3, we will see that the action of a free 
particle does not depend on $\k$, and $c$ is interpreted as the expectation value of 
photon's velocity. There is the deviation of the photon's velocity from the value $c$, 
because of the spacetime 
non-commutativity. The section 4 is devoted to the summary and the remarks.
\section{Review of $\k$-Minkowski spacetime}
In this section, we briefly review the $\k$-deformed Poincare algebra, and present 
discussions about velocities of particles in the $\k$-Minkowski spacetime.
\par    
Drinfeld \cite{Drinfeld:in} and Jimbo \cite{Jimbo:1985zk} have proposed a 
prescription how to define a quantum 
enveloping algebra, which is q-deformation of an universal enveloping algebra of a simple 
Lie algebra, in terms of a Hopf algebra.  Since the
Poincare algebra is not a simple Lie algebra, it is not straightforward to apply  
the Drinfeld-Jimbo prescription to the Poincare algebra. It is well known that the 
Poincare algebra can be obtained by contraction of the $3+2$ dimensional AdS algebra 
$o(3,2)$ which is a simple Lie algebra. In the 
limit $R \to \infty$ which is the AdS radius, the AdS algebra is reduced to the Poincare 
algebra. The contraction \cite{Celeghini:1990xx} $R\to \infty$ with 
$iR \log q \to \k^{-1}$ of the q-deformed AdS algebra gives a $\k$-deformed Poincare 
algebra
\cite{Lukierski:1991pn}. Here $\k$ is a real constant, and $q$ is an imaginary 
parameter\footnote{It is possible to use a real parameter 
\cite{Lukierski:1991ff,Lukierski:1992dt} instead 
of the imaginary $q$. In this case, contraction is $R \log q \to \k^{-1}$. } with 
$|q|=1$. The $\k$-deformation of the Poincare algebra is not unique 
\cite{Lukierski:1993wx,Magueijo:2001cr}, however which are 
related by 
redefinition of translation generators \cite{Kowalski-Glikman:2002we}. For example, the 
$\k$-Poincare algebra in the bicrossproduct basis \cite{Majid:1994cy} is given by  
\beeq
&&\left[p_{\m},p_{\n}\right]=0, \hsp1 \left[p_0,M_i\right] =0,  \hsp1
\left[p_i,M_j\right]= i\epsilon_{ijk}p_k         \label{2-3}    \\
&&\left[M_i,M_j\right]= i\epsilon_{ijk}M_k ,\hsp1 
\left[M_i,N_j\right]=i\epsilon_{ijk}N_k , \hsp1
\left[N_i,N_j\right]=i\epsilon_{ijk}M_k                     \\
&&\left[N_i,p_0\right]=ip_0  ,  \hsp1
\left[N_i,p_j\right] =i\d_{ij}\left(
\frac{\k}{2}(
1-e^{-2\frac{p_0}{\k}})
+\frac{1}{2\k}\mathbf{p}^2 \right)-\frac{i}{\k}p_ip_j   \label{2-4}
\eneq
with translation $p_{\m}$, rotation $M_i$ and boost $N_i$ generators. The $o(3)$ 
subalgebra is undeformed. The coproducts are given by 
\beeq
\D (M_i)\=M_i\otimes 1+1 \otimes M_i  \\
\D (N_i)\=N_i\otimes 1+\exp\left(-\frac{p_0}{\k}\right) \otimes N_i
+\frac{1}{\k}\epsilon_{ijk}p_j \otimes M_k   \\
\D (p_i)\=p_i\otimes 1+\exp\left(-\frac{p_0}{\k}\right) \otimes p_i   \label{2-6}  \\
\D (p_0)\=p_0\otimes 1+1 \otimes p_0  .
\eneq
When the coproducts are given, counits and antipodes are obtained uniquely. But we do not 
write the detailed form since we do not use them in the present paper. 
In the limit $\k \to \infty$, the Hopf algebra are reduced to the Poincare algebra. 
The mass shell condition is obtained as the Casimir operator;
\beeq
m^2=\left(
2 \k \sinh \left(\frac{p_0}{2\k} \right)\right)^2-\mathbf{p}^2e^{\frac{p_0}{\k}}.
                                        \label{2-1}
\eneq
The mass $m$ is not a physical one. A relation with a physical mass is suggested in 
\cite{Kowalski-Glikman:2002we}. We find that the expression gives the upper bound of the 
momentum; $p_ip^i\leq \k^2$. 
\par
It was noticed that quantum groups can be realized as 
symmetries in non-commutative spaces\footnote{As a review article, see \cite{wess}.} 
\cite{Manin:sz}. The $\k$-Poincare algebra is the symmetry 
in the $\k$-Minkowski spacetime \cite{Majid:1994cy} which is non-commutative;
\beeq
[X^i,X^0]=\frac{1}{\k}X^i, \hsp1 [X^i,X^j]=0. \label{2-2}
\eneq
Although there are some bases of the algebra 
\cite{Lukierski:1993wx,Magueijo:2001cr}, 
all of them correspond to the commutator \cite{Kowalski-Glikman:2002we}.
\par
In the commutative Minkowski spacetime, a  particle's velocity  is written as 
\beeq
v_i=\frac{\part E}{\part p^i}   \label{2-5}
\eneq
with $E=\sqrt{p^2+m^2}$. Many authors have assumed that this 
relation is true also in the $\k$-Minkowski spacetime \cite{Lukierski:1993wx}. For a 
massless particle, by solving (\ref{2-1}) and using the formula (\ref{2-5}) we have 
\beeqno
v^2=c^2 \exp \left(\frac{2E}{\k}\right).
\eneqno
So the velocity of a massless particle is deformed from the constant c, and has energy 
dependence. Because of the effect, the $\k$ has the constraint $|\k|^{-1}<10^{-33}$m from 
the astronomical observation of $\g$ rays \cite{Biller:1998hg}. On 
the other hand, in \cite{tamaki}, it was mentioned that the
rule for the differentiation with the momentum is changed since the momentum sum rule is 
$\k$-deformed. From the coproduct (\ref{2-6}), we notice that the sum of momenta 
$p_1$ and $p_2$ of two particles is deformed as
\beeqno
p_1+\exp \left(-\frac{E_1}{\k}\right) p_2
\eneqno
which is non-Abelian.
By applying this to the velocity formula, we have two kinds of velocities;
\beeqno 
V_{l}^i=\exp \left(-\frac{E}{\k}\right)V^i 
, \hsp1 V_{r}^i= \left(1+\frac{1}{\k}p\cdot V\right)^{-1}V^i.
\eneqno
We find that these two velocities satisfies $(V_{l}^i)^2=(V_{r}^i)^2=c^2$ 
for massless particles. 
\par
A common property with respect to them is that the velocities are determined uniquely 
when their energies are given. In 
other words, the particle velocities have no uncertainty. However, there have to be 
exist such an uncertainty, because time and spatial coordinates of particles cannot be 
observed simultaneously by the non-commutativity (\ref{2-2}). We would like to 
investigate the point in the next section. 
\section{Uncertainty of velocity}
Here we would like to see how velocities of point particles are deformed in the 
$\k$-Minkowski spacetime by using a method by which we can avoid arguments which depend 
on the bases of the $\k$-Poincare algebra. In order to do this, we consider a particle in 
coordinate space $X^{\m}$ instead of the momentum space.
\par
When a space is a non-commutative one, there is the possibility that a 
symmetry on the space becomes a quantum group \cite{Manin:sz}. The quantum group 
makes a commutator of coordinates covariant. In this formalism, a parameter 
of the non-commutativity is an invariant scale. By virtue of this, the 
doubly special relativity can be realized as a quantum group 
\cite{Kowalski-Glikman:2001gp}. We apply this to the $\k$-Minkowski spacetime (\ref{2-2}).
\par
All of known bases of the $\k$-Poincare algebra correspond to the commutator (\ref{2-2}) 
\cite{Kowalski-Glikman:2002we}. So we can perform an analysis which is independent on 
a choice of the basis if we start from the commutator (\ref{2-2}).
The Poincare quantum group transformation of the coordinates is defined by  
\beeqno
X'^{\m}=\L ^{\m}_{\ \n}X^{\n}+a^{\m}
\eneqno
with
\beeq
\eta_{\m\n}\L ^{\m}_{\ \r}\L ^{\n}_{\ \s}=\eta_{\r\s}.   \label{3-3}
\eneq
Here $\e^{\m \n}=(-,+,+,+)$ is the flat metric tensor.
We demand that the transformed coordinates $X'^{\m}$ also satisfy the commutator 
(\ref{2-2}). 
The Poincare quantum group does not determined only by the covariance 
of the commutator. Furthermore we demand that differentiation with respect to $X^{\m}$ 
is not deformed;
\beeqno
dX^{\m}\wedge dX^{\n}=-dX^{\n}\wedge dX^{\m}. 
\eneqno
In order for the condition (\ref{3-3}) to be compatible with the Poincare quantum group,
(\ref{3-3}) have to be commutative with all of the elements of the quantum group. By 
using the conditions, the Poincare quantum group \cite{zakrzewski} is determined uniquely;
\beeq
[\L ^{\m}_{\ \r},\L ^{\n}_{\ \s}]\=0 , \hsp1
[a^{\m},a^{\n}]= \frac{1}{\k}(\d^{\m}_i\d^{\n}_0-\d^{\n}_i\d^{\m}_0)a^i    \label{3-5}  \\
\left[a^{\m},\L ^{\n}_{\ k}\right]\= \frac{1}{\k}
[(\L^{\n}_{\ 0}-\d^{\n}_0)\L^{\m}_{\ \r}+(\L^{0}_{\ \r}-\d^0_{\r})\eta^{\m\n}]. 
                                                                          \label{3-6}
\eneq
which is the dual Hopf algebra \cite{Majid} of (\ref{2-3})-(\ref{2-4}).  
\par
Next we would like to see kinematics of a particle in the $\k$-Minkowski spacetime. The 
action of a free particle which is invariant under the Poincare quantum group (\ref{3-5})
and (\ref{3-6}) have the usual form; 
\beeq
S=-\fra12\int d \tau [\frac{1}{e}(\dot{X}^{\m})^2-em^2]  \label{3-2}
\eneq
with $\dot{X}^{\m}=\frac{d X^{\m}}{d \tau}$. On the other hand, when we consider the 
$\k$-deformed Poincare algebra (\ref{2-3})-(\ref{2-4}) as a bicrossproduct Hopf algebra 
\cite{Majid:1994cy}, the invariant metric is 
\beeqno
(X_0)^2-(X^i)^2+\frac{3}{\k}X_0
\eneqno
which is different from ours. It is not clear why they are different. The canonical 
Hamiltonian is also $\k$-independent
\beeqno
H=\frac{1}{2e}[P_{\m}P^{\m}-m^2].
\eneqno
The action of the free particle does not depend on the non-commutativity parameter $\k$ 
in the $\k$-Minkowski spacetime.  It is known that, in the non-commutative spacetime with 
$[X^{\m},X^{\n}]=i\theta^{\m \n}$, $\theta^{\m \n}$-dependent parts of free field 
theories' action 
vanish. This is the characteristic point in common. Although the action is 
$\k$-independent, the non-commutativity effects a change of physical quantities. In the 
rest of the present section, we see this explicitly.
\par
Let's consider a velocity of a particle which is described by the action 
(\ref{3-2}). Because of the spacetime 
non-commutativity, we cannot observe strict values of $X^i$ and $X^0$ simultaneously, 
and there exists an ambiguity of the particle's velocity. At first, we see the 
expectation value of the velocity. It is natural to make the expectation values of the 
coordinates $x^{\m}:=\langle X^{\m} \rangle$ be a 
solution of the equations of motion which are derived from the action (\ref{3-2}) with 
$X^{\m} \to x^{\m}$. The expectation value of the particle velocity is 
\beeqno
v^i\=\frac{d x^{i} }{d t }  = \frac{\dot{x}^i}{\dot{t}}.
\eneqno
By using the equation of motion, we have $v^2=c^2$ for massless particles. The expectation
value of the velocity satisfies the ordinary velocity formula. The invariant scale $c$ 
can be interpreted as the expectation value of a velocity of a massless particle.
\par
Next we would like to see a deviation from the expectation value. The expectation value 
of a deviation from the expectation value $x^{\m}$ satisfy the inequality 
\cite{Amelino-Camelia:1996gp}
\beeq
\langle (\D X^i)^2\rangle \langle (\D T)^2 \rangle \geq 
\frac{1}{4c^2 \k^2}|\langle  X^i \rangle|^2    \label{3-1}
\eneq
with $\D X^{\m}:=X^{\m}-x^{\m}$. Here the index $i$ is not summed.
The inequality means that time and spatial coordinates of a particle cannot be detected 
precisely 
at once except for at the origin in a coordinate system. When $|\langle  X^i \rangle|$ is 
large, so is the uncertainty. This depends which coordinate systems we use.   
\par
Let's consider a case in which a free particle starts from the origin 
$\langle  X^{\m} \rangle=0$ of the spacetime. At a later time, the particle is at 
$\langle  X^{\m} \rangle:=x^{\m}$ with $x^0 \neq 0$. 
Since the $o(3)$ subalgebra of the Poincare algebra is not 
$\k$-deformed, without loss of generality, we can choose a coordinate system with 
\beeq
L:=x^1\neq 0,  \hsp1 x^2=x^3=0 .         \label{3-7}
\eneq
In this case $\langle  X^i \rangle$ in the right hand side of (\ref{3-1}) is an 
expectation value of a propagation length of the particle during the time $x^0$. Hence 
when the length $L$ is very large so is the spacetime uncertainty. 
\par
Here we would like to see a deviation from the expectation value $v:=\langle V \rangle$. 
We denote $\d X^i=\sqrt{\langle (\D X^i)^2\rangle}$ and 
$\d T=\sqrt{\langle (\D T)^2\rangle}$.
Although the inequality (\ref{3-1}) does not forbid both of $\d X^i$ and $\d T$ to be 
large, by analogy with quantum mechanics, we assume that they are small with the 
equality (\ref{3-1}) satisfied. Under the 
condition with the configuration (\ref{3-7}), we find that the uncertainties along the 
directions $x^2$ and $x^3$ vanish; $\d X^2=\d X^3=0$, and $\d T$ and $\d X^1$ have  
non-zero values. The spacetime uncertainty brings 
about the following velocity's uncertainty;
\beeq
\d V^2 =  \mp 2 \frac{L^2}{t^3}\d T \pm 2 \frac{L}{t^2}\d X^1 
-4\frac{L}{t^3}\d X^1\d T \pm \frac{3L^2}{t^4}(\d T)^2+\frac{1}{t^2}(\d X^1)^2 
                                                            \label{3-4}
\eneq
up to second order in $\d X^i$ and $\d T$.
We assume that $\d X^i$ and $\d T$ are so small that the first two terms in (\ref{3-4}) 
are the dominant part. If we choose a configuration in which the first order terms are 
canceled 
with each other, this gives the minimum value of $\d V^2$ approximately. So the minimum 
value of $\d V^2$ is 
\beeq
\left. \d V^2 \right|_{\rm min}\approx -\frac{3v^3}{ |\k L|c}.        \label{3-8}
\eneq
Here we have used $v:= \frac{L}{t}$.
\section{Summary and remarks}
In this paper, we have investigated free particle's motion in the $\k$-Minkowski 
spacetime in which the Poincare group becomes the quantum group. The particle's action 
is not $\k$-deformed, since the invariant metric of the $\k$-Minkowski spacetime does 
not have  $\k$-dependence in our formalism. When we 
define particle's velocity, there exists uncertainty of the velocity because the 
spacetime coordinates are non-commutative operators with the commutator (\ref{2-2}). 
The point is different from the 
discussions in momentum spaces \cite{Lukierski:1993wx,tamaki}. We saw that the 
particle's velocity can depend on $\k$, even if the action is $\k$-independent. 
\par
In ordinary relativistic classical theories, massless particles propagate only with 
the velocity $c$ which is defined by $c=299792458m/s$. However in the $\k$-Minkowski 
spacetime, the massless particle's velocity have the uncertainty   
\beeqno
\left. \d V^2\right|_{\rm min} \approx -\frac{3c^2}{ |\k L|}
\eneqno
which does not depend on the energy of the particle but on the propagation length $L$ of 
the particle. When $L$ is small, so is the spacetime uncertainty 
(\ref{3-1}). However for a small $L$, the velocity's uncertainty is large.  
The constant $c$ can be interpreted as the expectation value of the photon's velocity.
\par 
In order to compatible with the observed value of $c$, we have the bound 
$|L \k|>10^{20}$. 
The typical length of light paths in experiments of velocity of light is of order 
$10^2$-1$0^{3}$m \cite{froome}. 
So $\k$ is $|\k|>10^{17}$m which is weaker restriction than the one which is derived by 
using the formula (\ref{2-5}).
\par
In a non-relativistic limit $\frac{v}{c}\to 0$, the uncertainty (\ref{3-8}) of the 
velocity can be negligible. The result is compatible with the fact that the only Lorentz 
boost part of the Poincare algebra is $\k$-deformed.

\vspace{10pt}

{\large \bf Acknowledgment} \\

\vspace{4pt}

The author would like to thank the Yukawa institute for hospitality, where a part of the 
research was done.


\begin{thebibliography}{10}

\bibitem{Takeda:1998ps}
M.~Takeda {\it et al.},
``Extension of the cosmic-ray energy spectrum beyond the predicted  
Greisen-Zatsepin-Kuzmin cutoff'',
Phys.\ Rev.\ Lett.\  {\bf 81}, 1163 (1998), astro-ph/9807193.

\bibitem{greisen}
K.~Greisen, ``End to the Cosmic-ray spectrum?'', Phys.\ Rev.\ Lett.\  {\bf 16}, 748 
(1966); G.~T.~ Zatsepin and V.~A.~Kuzmin, 
``Upper limit of the spectrum of cosmic rays'', Sov.~Phys.~JETP Lett.~{\bf 4}, 78 (1966).

\bibitem{Quinn:dj}
 M.~Punch {\it et al.},
``Detection Of Tev Photons From The Active Galaxy Markarian 421'',
Nature {\bf 358}, 477 (1992); J.~Quinn {\it et al.},
``Detection Of Gamma-Rays With E$>$300-Gev From Markarian 501'',
Astrophys.\ J.\  {\bf 456}, L83 (1996); M.~Catanese {\it et al.},
``Discovery of Gamma-Ray Emission above 350 GeV from the BL Lacertae Object 1ES 
2344+514'', Astrophys.\ J.\  {\bf 501}, 616 (1998).

\bibitem{sato}
H.~Sato and T.~Tati, ``Hot Universe, Cosmic Rays of Ultrahigh Energy and Absolute 
Reference System'', Prog.~Theor.~Phys.~{\bf 47}, 1788 (1972).

\bibitem{Coleman:1998ti}
T.~Kifune, ``Invariance violation extends the cosmic ray horizon?''
Astrophys.\ J.\  {\bf 518}, L21 (1999), astro-ph/9904164;
S.~R.~Coleman and S.~L.~Glashow, ``High-energy tests of Lorentz invariance'',
Phys.\ Rev.\ D {\bf 59}, 116008 (1999), hep-ph/9812418;
G.~Amelino-Camelia and T.~Piran,
``Planck-scale deformation of Lorentz symmetry as a solution to the UHECR  and the 
TeV-gamma paradoxes'', Phys.\ Rev.\ D {\bf 64}, 036005 (2001), astro-ph/0008107.

\bibitem{tamaki}
T. Tamaki, T. Harada, U. Miyamoto and T. Torii, "Have we already detected 
astrophysical symptoms of space-time noncommutativity?", gr-qc/0111056;    
``Particle velocity in noncommutative space-time'', Phys.\ Rev.\ D {\bf 66}, 105003 
(2002), gr-qc/0208002.

\bibitem{Amelino-Camelia:2000mn}
G.~Amelino-Camelia, ``Relativity in space-times with short-distance structure governed 
by an observer-independent (Planckian) length scale'',
Int.\ J.\ Mod.\ Phys.\ D {\bf 11}, 35 (2002), gr-qc/0012051;
``Testable scenario for relativity with minimum-length,''
Phys.\ Lett.\ B {\bf 510}, 255 (2001), hep-th/0012238.

\bibitem{Kowalski-Glikman:2001gp}
J.~Kowalski-Glikman, ``Observer independent quantum of mass'',
Phys.\ Lett.\ A {\bf 286}, 391 (2001), hep-th/0102098; 
N.~R.~Bruno, G.~Amelino-Camelia and J.~Kowalski-Glikman,
``Deformed boost transformations that saturate at the Planck scale'',
Phys.\ Lett.\ B {\bf 522}, 133 (2001), hep-th/0107039.


\bibitem{Lukierski:1991pn}
J.~Lukierski, H.~Ruegg, A.~Nowicki and V.~N.~Tolstoi,
``Q deformation of Poincare algebra'', Phys.\ Lett.\ B {\bf 264}, 331 (1991).

\bibitem{Majid:1994cy}
S.~Majid and H.~Ruegg,
``Bicrossproduct structure of kappa Poincare group and noncommutative geometry'',
Phys.\ Lett.\ B {\bf 334}, 348 (1994), hep-th/9405107.

\bibitem{Lukierski:1993wx}
J.~Lukierski, H.~Ruegg and W.~J.~Zakrzewski,
``Classical and quantum mechanics of free kappa relativistic systems'',
Annals Phys.\  {\bf 243}, 90 (1995), hep-th/9312153.




\bibitem{Magueijo:2001cr}
J.~Magueijo and L.~Smolin,
``Lorentz invariance with an invariant energy scale'',
Phys.\ Rev.\ Lett.\  {\bf 88}, 190403 (2002), hep-th/0112090;
P.~Kosinski, J.~Lukierski, P.~Maslanka and J.~Sobczyk,
``The Classical basis for kappa deformed Poincare (super)algebra and the second kappa 
deformed supersymmetric Casimir'',
Mod.\ Phys.\ Lett.\ A {\bf 10}, 2599 (1995), hep-th/9412114;
P.~Maslanka, ``Deformation map and Hermitian representations of $\k$-Poincare algebra'', 
J.\ Math.\ Phys.\ {\bf 34}, 6025, (1993).

\bibitem{Kowalski-Glikman:2002we}
J.~Kowalski-Glikman and S.~Nowak,
``Doubly special relativity theories as different bases of kappa-Poincare  algebra'',
Phys.\ Lett.\ B {\bf 539}, 126 (2002), hep-th/0203040.


\bibitem{Drinfeld:in}
V.~G.~Drinfeld,
``Quantum Groups'', J.\ Sov.\ Math.\  {\bf 41}, 898 (1988)
[Zap.\ Nauchn.\ Semin.\  {\bf 155}, 18 (1986)].

\bibitem{Jimbo:1985zk}
M.~Jimbo,
``A Q Difference Analog Of U(G) And The Yang-Baxter Equation'',
Lett.\ Math.\ Phys.\  {\bf 10}, 63 (1985);
``A Q Analog Of U (Gl (N+1)), Hecke Algebra And The Yang-Baxter Equation'',
Lett.\ Math.\ Phys.\  {\bf 11}, 247 (1986).

\bibitem{Celeghini:1990xx}
E.~Celeghini, R.~Giachetti, E.~Sorace and M.~Tarlini,
``The Three-Dimensional Euclidean Quantum Group E(3)-Q And Its R Matrix,''
J.\ Math.\ Phys.\  {\bf 32}, 1159 (1991).

\bibitem{Lukierski:1991ff}
J.~Lukierski, A.~Nowicki and H.~Ruegg,
``Real forms of complex quantum anti-De Sitter algebra U-q(Sp(4:C)) and their 
contraction schemes'', Phys.\ Lett.\ B {\bf 271}, 321 (1991), hep-th/9108018.


\bibitem{Lukierski:1992dt}
J.~Lukierski, A.~Nowicki and H.~Ruegg,
``New quantum Poincare algebra and k deformed field theory'',
Phys.\ Lett.\ B {\bf 293}, 344 (1992).


\bibitem{wess} 
J.~Wess, ``q-Deformed Heisenberg Algebras'', math-ph/9910013.
 
\bibitem{Manin:sz}
S.~L.~Woronowicz,
``Compact Matrix Pseudogroups'', Commun.\ Math.\ Phys.\  {\bf 111}, 613 (1987);
``Differential calculus on compact matrix pseudogroups (quantum groups)'', 
Commun.\ Math.\ Phys.  {\bf 122}, 125 (1989);
Y.~I.~Manin,
``Quantum groups and non-commutative geometry'', 
Centre de recherches mathematiques (1989);
``Multiparametric Quantum Deformation Of The General Linear Supergroup'',
Commun.\ Math.\ Phys.\  {\bf 123}, 163 (1989).


\bibitem{Biller:1998hg}
G.~Amelino-Camelia, J.~R.~Ellis, N.~E.~Mavromatos, D.~V.~Nanopoulos and S.~Sarkar,
``Potential Sensitivity of Gamma-Ray Burster Observations to Wave Dispersion in Vacuo'',
Nature {\bf 393}, 763 (1998), astro-ph/9712103;
S.~D.~Biller {\it et al.},
``Limits to quantum gravity effects from observations of TeV flares in  active galaxies,''
Phys.\ Rev.\ Lett.\  {\bf 83}, 2108 (1999), gr-qc/9810044;
R.~Gambini and J.~Pullin,
``Nonstandard optics from quantum spacetime'',
Phys.\ Rev.\ D {\bf 59}, 124021 (1999), gr-qc/9809038;
G.~Amelino-Camelia, J.~Lukierski and A.~Nowicki,
``Distance measurement and kappa-deformed propagation of light and heavy  probes'',
Int.\ J.\ Mod.\ Phys.\ A {\bf 14}, 4575 (1999) gr-qc/9903066;
G.~Amelino-Camelia and S.~Majid,
``Waves on noncommutative spacetime and gamma-ray bursts'',
Int.\ J.\ Mod.\ Phys.\ A {\bf 15}, 4301 (2000), hep-th/9907110.

\bibitem{zakrzewski}
S.~Zakrzewski, ``Quantum Poincare group related to the kappa -Poincare algebra'',
J.~Phys.~{\bf A27} 2075 (1994).


\bibitem{Majid}
S.~Majid, ``Foundations of Quantum Group Theory'', Cambridge University Press (1995).



\bibitem{Amelino-Camelia:1996gp}
G.~Amelino-Camelia,
``Enlarged bound on the measurability of distances and quantum  kappa-Poincare group'',
Phys.\ Lett.\ B {\bf 392}, 283 (1997), gr-qc/9611016.
 
\bibitem{Giller:xg}
S.~Giller, P.~Kosinski, M.~Majewski, P.~Maslanka and J.~Kunz,
``More About Q Deformed Poincare Algebra'',
Phys.\ Lett.\ B {\bf 286}, 57 (1992).

\bibitem{froome}
K.~D.~Froome and L.~Essen, ``The Velocity of Light and Radio Waves, Academic, New York    
(1969).

\end{thebibliography}
\end{document}